\documentclass[a4paper]{paperStyle}
\usepackage{epsfig, isolatin1}

\begin{document}

\title{Warm molecular gas in the Galactic center region}

\author{ N.\,J. Rodr\'{\i}guez-Fern\'andez \and J. Mart\'{\i}n-Pintado
\and A. Fuente \and P. de Vicente}

\institute{
Observatorio Astron\'omico Nacional,
Apdo. 1143, E28800 Alcal\'a de Henares, Spain}

\maketitle 

\begin{abstract}
We present some recent studies on the physical conditions and the chemistry
of the interstellar medium (ISM) in the Galactic center region (GCR).
Large scale shocks in the context of a bar potential can play a role in explaining
the observed properties, at least in some particular regions.
Alternatively, we propose that the observed
properties (inhomogeneity, high temperatures of the molecular
gas, rich chemistry,...)
of the GCR ISM can be the consequence of the recent past ($10^6-10^7$ years ago)
of massive star formation in the GCR.


\end{abstract}

%\section{Introduction}
The clouds of the Central Molecular Zone (CMZ; $R<250$~pc) in the
Galactic center region (GCR),
are more turbulent  and  denser
than the clouds of the  Galactic disk.
In the CMZ there is  a widespread warm gas component with temperatures
of $\sim$ 150~K but without associated warm dust
(the dust temperature is not higher than 40~K).
In the disk of the Galaxy, those high gas temperatures are only found
in small regions where the dust is also warm.
In the CMZ it is also possible to detect (and with
high abundances) widespread emission of molecules that in the rest of the
Galaxy are only present in small {\it hot cores} or shocked regions.
The origin of the high gas temperatures and the rich chemistry of the CMZ
clouds is still not clear.

 Rodr{\' \i}guez-Fern\'andez et al. (2001)
 have presented the first direct estimation of the total
 column density of warm gas in the CMZ clouds thanks to
 {\it Infrared Space Observatory} (ISO) observations of
 the lowest H$_2$ pure-rotational lines.
 The column density of warm gas ($\sim 150$ K) is
 typically 1--2 10$^{22}$ cm$^{-2}$. On average,
 the warm H$_2$ column densities represent a fraction of 30 $\%$ of the gas
 traced by CO. It is not easy to explain the heating of such quantities
 of warm gas.
 The H$_2$ pure-rotational lines
 can be explained both as arising in low velocity shocks ($\sim 10$ km\,s$^{-1}$)or PDRs (Rodr{\' \i}guez-Fern\'andez et al. 2001a).
 However, to explain the total amount of warm gas measured in the CMZ clouds
 it is needed to invoke several of these shocks or PDRs (or both) in the
 line of sight.

On the one hand,
the high abundances in gas phase of molecules linked to the grain chemistry
like SiO, C$_2$H$_5$OH or NH$_3$ (Mart{\' \i}n-Pintado et al. 1997, 2001;
Rodr{\' \i}guez-Fern\'andez et al. 2001)
point to shocks
as the exciting mechanism since the dust temperature is too low to evaporate grain
mantles and in addition
these are fragile molecules that are easily dissociated by UV radiation.
The high abundance of these molecules are difficult to explain in a photo-dissociation
region (PDR) even invoking a 100 km\,s$^{-1}$ turbulence
(Rodr{\' \i}guez-Fern\'andez 2002).
The large line-widths of the lines also point to shocks (Wilson et al. 1982).
Large scale shocks due to cloud collisions in the context of a bar potential
have been invoked to explain the high SiO abundances,
the non-equilibrium H$_2$ ortho-to-para ratio, and the high
fraction of warm H$_2$ measured in the clouds located at the extremes
of the CMZ  ($l\sim 1.5^\circ$ and
$l\sim -1^\circ$; Rodr{\' \i}guez-Fern\'andez et al. 2001a;
H\"uttemeister et al. 1998).

On the other hand,
ISO has detected fine structure lines of ions like Ne~{\sc ii},
S~{\sc iii} or N~{\sc ii} toward many ``molecular'' clouds
(Rodr{\' \i}guez-Fern\'andez et al. 2001b, 2002). These lines
should arise in H~{\sc ii} regions and there must be a
photo-dissociation region (PDR) in the interface with the cold neutral gas.
This is the case in the Radio Arc region were the Quintuplet
and the Arches clusters
ionize a large region with a radius of more than 15 pc (Rodríguez-Fernández
et al. 2001b).
Furthermore, taking into account the effective temperature of the
radiation and the total number of Lyman continuum photons emitted by the
clusters (estimated independently from their stellar content and from the
ionized ISM in their surroundings) it is easy to derive a far-ultraviolet
(FUV) flux of 10$^4$ in units of the field of Habing (1968)
(see Rodríguez-Fernández 2002). In perfect agreement with
the FUV field suggested by
the comparison of the H$_2$ pure-rotational lines and the models
(Rodríguez-Fernández et al. 2001a).
Therefore, we conclude that at least for some of the clouds observed by ISO,
the warm H$_2$ arises in PDRs.
(It is important to remark that in PDRs of moderate density 
($\sim 10^{3}$\,cm$^{-3}$) the gas can be heated to
temperatures of $\sim 200$ K without heating the dust to temperatures higher
than $\sim 35$~K; Hollenbach et al. 1991).

 The importance of X-radiation in the Galactic center region 
 should also be revised.
 The extended emission of the 6.4 keV fluorescent line of neutral or low ionized
 Fe reveals the interaction of hard X-rays with
 neutral or low ionized  gas (Koyama et al. 1996).
 The striking correlation of the Fe 6.4 keV line and the SiO
 emission extend that interaction to the cold neutral gas, at least
 indirectly (Mart{\' \i}n-Pintado et al. 2000).

The study of the warm molecular gas with high resolution have given us new
hints on the possible heating mechanisms and the origin of the chemistry.
The NH$_3$ VLA observations of the envelope of Sgr B2 have shown that
the distribution of the warm gas is very inhomogeneous
(Mart{\' \i}n-Pintado et al. 1999).
It is dominated by expanding ($\sim 10$~km$^{-1}$)
shells probably due to the interaction of 
Wolf-Rayet winds with the surrounding medium.
This scenario of a cluster of massive stars (including evolved massive
stars) can explain many of
the known characteristics of the Sgr B2 envelope:
$\sim 10$~km$^{-1}$ turbulence and shock chemistry,  warm gas temperatures,
the correlation of the Fe 6.4 keV and the SiO lines,...
In this context, shocks, PDRs, and X-rays would be present.
For instance, with two
shells in the line of sight there would be 4 PDRs and 4 low velocity shocks
and the total column density of warm gas would be $\sim 10^{22}$~cm$^{-2}$
as observed.

 As previously mentioned, large scale shocks
 in the context of a bar potential could
 dominate the heating and
 the chemistry of the clouds located near the Inner Lindblad Resonances
 (presumably around $l \sim 1.5^\circ$ and $l \sim - 1^\circ$).
 In these regions cloud collisions are expected due to the presence of
 self-intersecting  x1 orbits and the intersection of x1 and x2 orbits
 (see Binney et al. 1991).

  However,
  the large scale properties of the CMZ are similar to those of
  the Sgr B2 envelope. The scenario of a very inhomogeneous medium
  created by evolved massive stars could also apply for the CMZ as a whole.
  Indeed, there are several evidences that there has been  a burst of massive star formation in the GCR $\sim 10^{6-7}$ yr ago (see Morris \& Serabyn 1996).
  The well known Quintuplet and Arches clusters
  as well as the proposed cluster in the envelope of Sgr B2
  should have been formed in that burst.
  We suggest that the physical properties (turbulence, heating, inhomogeneity...)
  and the high abundance of refractory or complex organic molecules
  in the GCR clouds can be
  explained in the context of that burst of massive  star formation.
  Thus, the present state of the GCR ISM can be the consequence of
  the recent past of massive star formation in the central hundreds parsecs
  of the Galaxy.

\begin{acknowledgements}

NJR-F acknowledges a predoctoral fellowship from
{\em Consejer\'{\i}a de Educaci\'on de la Comunidad de Madrid}

\end{acknowledgements}

\end{document}